\documentclass[intlimits,twoside,a4paper]{article}

\usepackage[cp1251]{inputenc}
\usepackage{xcolor}
\usepackage[eqsecnum]{cmpj3}

\usepackage{bm}
\issue{2020}{23}{2}{23701}
\doinumber{10.5488/CMP.23.23701}

\title[On the energy of topological defect lattices]%
{On the energy of topological defect lattices}
\author[B. Berche,  S. Fumeron, F. Moraes]{B. Berche\refaddr{label1},  S. Fumeron\refaddr{label1}, F. Moraes\refaddr{label2}}
\addresses{
\addr{label1} Dynamique et Sym\'etries, Laboratoire de Physique et Chimie Th\'eoriques,  CNRS - Universit\'e de Lorraine, UMR 7019,
    54506 Vand\oe uvre les Nancy, France
\addr{label2} Departamento de F\'{\i}sica, Universidade Federal Rural de Pernambuco, 
 52171--900 Recife, PE, Brazil
}
\date{Received January 28, 2020, in final form February 27, 2020}
\begin{document}
\maketitle

\begin{abstract}
Since the logarithm function is the solution of Poisson's equation in two dimensions, it appears as the Coulomb interaction in two dimensions, the interaction between  Abrikosov flux lines in a type II superconductor, or between line defects in elastic media, and so on. Lattices of lines interacting logarithmically are, therefore, a subject of intense research due to their manifold applications. The solution of the Poisson equation for such lattices is known in the form of an infinite sum since the late 1990's. In this article we present an alternative analytical solution, in closed form, in terms of the Jacobi theta function. 
\keywords topological defect, cosmic string, flux line 

\end{abstract}

\hspace*{2.7em}{\em A paper dedicated to our friend Ihor Mryglod}

\hspace*{2.7em}{\em on the occasion of his 60th birthday.}

\date{\today}

\section{Introduction}\label{sec1}
As intriguing as topological defects might be, crystals made out of them appear to be even more exotic, like the soliton lattice that forms in doped polyacetylene \cite{chen1985soliton}, for instance. Examples of topological defect lattices abound in Condensed Matter Physics where one might find lattices of parallel screw dislocations in solids \cite{kosevich2004oscillations}, vortex lattices in rotating superfluids \cite{lounasmaa1999vortices} and in Bose-Einstein condensates~\cite{adhikari2019vortex}, as well as the much studied magnetic flux lattices in type II superconductors \cite{abrikosov1957magnetic}. Liquid crystals contribute with lattices of disclinations in nematics \cite{murray2014creating}, and with lattices of screw dislocations in cholesterics (known as twist grain boundaries) \cite{renn1988abrikosov}. Nevertheless, the most fascinating topological defect lattices are found in the realm   of chiral  liquid crystals \cite{bahr2001chirality} where   skyrmions \cite{fukuda2011quasi}, hopfions (3D skyrmions) \cite{ackerman2015self}, merons (half-skyrmions)~\cite{duzgun2018comparing} and even knots \cite{tai2019three} may  form regular arrays. Off  this planet one might have magnetic flux tube lattices in  neutron stars \cite{mazur2015casimir} and crystals of cosmic strings or of cosmic domain walls, which have  been considered as possible candidates for solid dark matter models \cite{bucher1999dark,battye2006elastic}. All this zoo of topological defects shares a common origin: phase transitions involving break of symmetry. Not surprisingly thus, the Kibble-Zurek mechanism \cite{kibble1976topology,zurek1985cosmological} of defect formation   applies both to cosmic strings and disclinations in nematic liquid crystals \cite{Bowick1994}. A common feature of most of the aforementioned topological defect crystals is a logarithmic interaction of line defects in the lattice, for large enough separation between them, so the defect internal structure may be neglected. This leads to the problem of performing infinite log sums, much tackled in the 1990's. 

An important step was done in the calculation of energies and forces between particles interacting logarithmically by \cite{gronbech1996summation,gronbech1999summation} which much improved the efficiency of computer simulations. The expressions were obtained in terms of products of elementary trigonometric or hyperbolic functions.  
In this work we move a step forward and obtain for the solution of Poisson equation a closed form for the logarithmic sum involving Jacobi theta functions. These functions are special functions of complex variables which appear in the theory of elliptic functions which are ubiquitous in mathematical physics.

Logarithmic potential appears as a solution of the two-dimensional Poisson equation, and thus, describes the interaction  in a two-dimensional Coulomb gas, so there is   no mystery in the appearance of elliptic and Jacobi theta functions in the vortices-driven Berezinskii-Kosterlitz-Thouless transition~\cite{Berezinskii71,KosterlitzThouless73,Kosterlitz74,Kosterlitz2016}.  
Indeed, applying the theory of conformal mappings which hold at the critical point of second order phase transitions in two-dimensional systems,  these special functions enable us to obtain a closed expression for the correlation functions in $XY$ models \cite{berche2003bulk}. More generically, they appear in the Schwarz-Christoffel mapping and related conformal mappings in the complex plane \cite{LavrentievChabat,diFrancescoMathieuSenechal97,TalapovEtAl,PhysRevE.60.3853}. 

We consider an infinite lattice of parallel string-like defects in 3D or, equivalently, point-like defects in 2D, interacting logarithmically. Our interest is to find the  energy and, consequently, the force on  a test defect  due to its interaction with the lattice. Although previous results for finite lattices with periodic boundary conditions have been reported \cite{glasser1974evaluation,stremler2004evaluation,tyagi2004effective,tyagi2006logarithmic}, to the best of our knowledge, this is the first time where a closed form for the logarithmic sum is achieved for the infinite lattice.

\subsection{Rectangular lattice }

Let us consider a rectangular Bravais lattice in $\mathbb{R}^2$ generated by the basis vectors $\vec{a}_1 =a\hat{x}$ and $\vec{a}_2 = b\hat{y}$ such that a  point of the lattice located at $\vec{R}_{mn}=m\vec{a}_1+n\vec{a}_2$ is  associated to the pair $(m,n)\in \mathbb{Z}^2$. To each point of the lattice we associate a defect. It is our purpose to find the potential due to this array of defects, assuming the superposition principle. 
That is, we want to perform the sum
\begin{eqnarray}
V(\vec{r})= \lambda\sum_{(m,n)\in\mathbb{Z}^2} \ln |\vec{r} -\vec{R}_{mn}|^2 , \label{fv} \\
\nonumber
\end{eqnarray}
where $\vec{r}=x\hat{x}+y\hat{y}$ is the  position of a test defect and $\lambda$ is the ``charge'' of the logarithmic interaction. 
Obviously, the function defined by equation~(\ref{fv}) is a solution of the 2D Poisson equation
\begin{equation}
(\partial^2_{xx}  +\partial^2_{yy})V = 2\piup\lambda  \sum_{(m,n)\in\mathbb{Z}^2} \delta \left( x- ma \right)\delta \left( y- nb \right) \label{laplacian}
\end{equation}
and, therefore, we will not be concerned with additive constants appearing in the logarithmic sum. This is the essence of the regularization process that we need to use since the ``raw'' sum in equation~(\ref{fv}) naturally diverges. 

Now, defining
\begin{align}
\varphi &= x+\ri y , &  \bar{\varphi} &= x-\ri y \label{phi}, \\
\sigma&= ma +  \ri nb, &  \bar{\sigma}&=ma  -\ri nb
\end{align} 
 we write equation~(\ref{fv}) as
\begin{equation}
V(x,y)= \lambda\sum_{(m,n)\in\mathbb{Z}^2} \ln \left[ (\varphi - \sigma)(\bar{\varphi} - \bar{\sigma}) \right] . \label{fvnew} 
\end{equation}
As mentioned above, the sums in equations~(\ref{fv}) and (\ref{fvnew}) diverge but can be regularized by subtracting  constant divergent terms
as we will see below. 

Choosing to first perform the sum over $n$ in equation~(\ref{fvnew}), we have
\begin{eqnarray}
V(x,y) &=& \lambda\sum_{m=-\infty}^{\infty}  \ln \left[ (\varphi -ma)(\bar{\varphi}-ma) \right]  \nonumber \\ 
&+& \lambda\sum_{m=-\infty}^{\infty}  \sum_{n=1}^{\infty} \ln\left\lbrace  [(\varphi - ma)^2  +n^2 b^2  ] 
 [(\bar{\varphi} - ma)^2 +n^2 b^2  ]\right\rbrace   . \label{sum1}
\end{eqnarray}
Note that, in changing the sum over $\mathbb{Z}^-$ into a sum  over $\mathbb{Z}^+$,   we have   $\sigma \rightarrow \bar{\sigma}$   such that $(\varphi - \sigma)(\bar{\varphi}-\bar{\sigma})$ from the $\mathbb{Z}^-$ sum becomes $(\varphi - \bar{\sigma})(\bar{\varphi}-\sigma)$.

Now, using the identity
\begin{equation}
\prod_{n=1}^{\infty} \left( 1 + \frac{z}{n^2 A + B} \right) = \sqrt{\frac{B}{B+z}} \frac{\sinh ( \piup\sqrt{B+z}/\sqrt{A})}{\sinh (\piup \sqrt{B/A})}\, ,
\end{equation}
 equation~(\ref{sum1}) writes
\begin{align}
  V(x,y)  &=  \lambda\sum_{(m,n)\in\mathbb{Z}^2} \ln  \left( m^2a^2+ n^2b^2 \right) + \lambda\sum_{m=-\infty}^{\infty} \ln \left( m^2 a^2 \right) \nonumber \\
 & +\lambda\sum_{m=-\infty}^{\infty} \ln \left\lbrace \frac{ \sinh[\piup (\varphi -ma) /b]}{\sinh (\piup ma/b)}  \right\rbrace  
+\lambda\sum_{m=-\infty}^{\infty} \ln \left\lbrace  \frac{ \sinh[\piup(\bar{\varphi} -ma) /b]} {\sinh (\piup ma/b)} \right\rbrace , 
\label{sum2}
\end{align}
 which results in
\begin{align}
V(x,y)  & =   \lambda\sum_{(m,n)\in\mathbb{Z}^2} \ln  \left( m^2a^2+ n^2b^2 \right) + 2 \lambda\sum_{m=1}^{\infty} \ln \left( m^2 a^2 \right)+\lambda\ln\left( b^2 / \piup^2\right) \nonumber \\
 & +   \lambda\ln[ \sinh(\piup\varphi /b) \sinh(\piup\bar{\varphi} /b)] \nonumber \\
&+ \lambda\sum_{m=1}^{\infty} \ln \left\lbrace \frac{  \sinh[\piup (\varphi -ma) /b] \sinh[\piup (\varphi +ma) /b]}{\sinh^2 (\piup ma/b)} \right\rbrace \nonumber \\
 & +   \lambda\sum_{m=1}^{\infty} \ln \left\lbrace  \frac{ \sinh[\piup (\bar{\varphi} -ma) /b] \sinh[\piup (\bar{\varphi} +ma) /b]}{\sinh^2 (\piup ma/b)}  \right\rbrace . 
 \label{sum3}
\end{align}
The first three terms in equation~(\ref{sum3}) just add up to an infinite constant and can be removed from the potential since it will still be a solution of equation~(\ref{laplacian}).

The remaining terms can be evaluated by the use of the identity
\begin{equation}
\prod_{m=1}^{\infty} \left[ \cos^2 z + \sin^2 z \coth^2 (m\piup\chi) \right] =  \csc z \ \frac{\vartheta_ 1 (z,\re^{-\piup\chi})}{\vartheta_1'(0,\re^{-\piup\chi})}\,,
\end{equation}
in terms of the Jacobi theta function $\vartheta_1$ and its first derivative w.r.t. $z$.  As warned by Abramowitz and Stegun \cite{abramowitz1964handbook}, there is a bewildering variety of notations for the theta functions. The one we use here, given by \cite{abramowitz1964handbook} and \cite{olver2010nist}, is such that they have the following Fourier representation:
\begin{eqnarray}
\vartheta_1 (z, \re^{-\piup\chi})= 2\sum_{n=0}^ {\infty} (-1)^n \re^{-\piup\chi (n+1/2)^2} \sin \left[(2n+1 )z\right], \label{Fourier}
\end{eqnarray}
where $z$ and $\chi$ are complex numbers.
With the help of the above relations, equation~(\ref{sum3}) and, therefore, the regularized equation~(\ref{fv}) takes a surprisingly simple form
\begin{equation}
V(x,y) = \lambda \ln \left[\frac{-|\vartheta_1 (\ri\piup \varphi /b, \re^{-\piup a/b})|^2 }{\sin(\ri\piup\varphi/b)\sin(\ri\piup\bar{\varphi}/b)}  \right] - 2\lambda\ln \left[ \vartheta_1'(0,\re^{-\piup a/b})\right] . \label{Vreg}
\end{equation}
    This results in the following expression  in terms of the coordinates $x$ and $y$:
\begin{equation}
V(x,y) = \lambda \ln(x^2 + y^2) + \lambda  \ln \left[\frac{|\vartheta_1 (\frac{\piup}{b}(\ri x-y), \re^{-\piup a/b})|^2}{\cosh^2(\piup x/b)-\cos^2(\piup y/b)}  \right] - 2\lambda\ln \left[ \vartheta_1'(0,\re^{-\piup a/b})\right] . \label{Vregxy}
\end{equation}


 Equation (\ref{Vregxy}) must have logarithmic singularities at the defect sites since it is a compact version of equation~(\ref{fv}). Nevertheless, it seems to have extra singularities at $(x,y)=(0,nb)$. Since both $\cos (\piup y/b) $ and the $\vartheta$ function are periodic in $y$ with periodicity $b$, it suffices to examine this question near $(x,y)=(0,0)$. A closer look at equation~(\ref{Vreg}) indicates that there is no extra singularity there since  $\vartheta_1 (0, \re^{-\piup\chi})=0 $ and
\begin{equation}
     \lim_{z\rightarrow 0} \frac{ \vartheta_1 (z, \re^{-\piup\chi})}{\sin z} = \lim_{z\rightarrow 0} \frac{ \vartheta_1' (z,\re^{-\piup\chi})}{\cos z}= \vartheta_1'(0,\re^{-\piup\chi}) =  2\sum_{n=0}^ {\infty} (-1)^n \re^{-\piup\chi (n+1/2)^2} (2n+1 )\,, \label{lim}
\end{equation}
where the values of $\vartheta_1(0,q)$ and $\vartheta_1'(0,q)$ were obtained from equation~(\ref{Fourier}).
Hence,  $V(0,0)= \lambda \ln(x^2 + y^2)|_{(x,y)=(0,0)}$ as it should. Likewise, by changing the origin in  equation~(\ref{Vregxy}) to $(0,nb)$, it follows that $V(0,nb)= \lambda \ln[x^2 + (y-nb)^2]|_{(x,y)=(0,nb)}$. 

\begin{figure}[!t]
\includegraphics[scale =0.4]{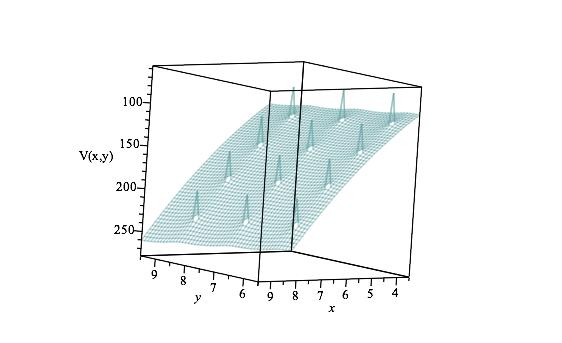}
\includegraphics[scale =0.4]{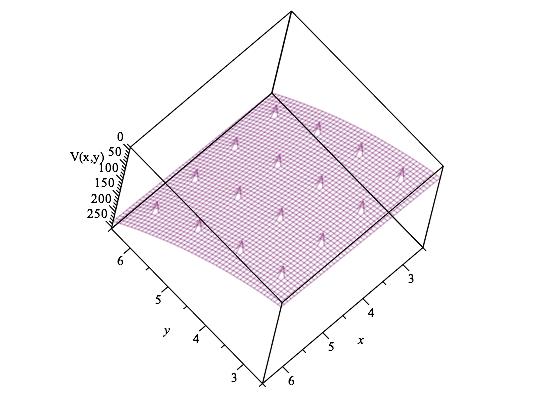}
\caption{(Colour online) Three-dimensional plot of the ``potential function'' $ V(x,y)$ for the rectangular and square lattices, respectively.}
	\label{fig_Berche1}
\end{figure}

In equation~(\ref{lim}) we see that, since $\chi$ is real,     $\vartheta_1'(0,\re^{-\piup\chi})$ is also real and, therefore, $[\vartheta_1'(0,\re^{-\piup\chi})]^2 = |\vartheta_1'(0,\re^{-\piup\chi})|^2$. This way, we rewrite equation~(\ref{Vregxy}) as 
\begin{equation}
V(x,y) = \lambda \ln \left[ \frac{x^2 + y^2}{\cosh^2(\piup x/b)-\cos^2(\piup y/b)} \right] + \lambda  \ln \left|\frac{\vartheta_1 (\frac{\piup}{b}(\ri x-y), \re^{-\piup a/b})}{\vartheta_1'(0,\re^{-\piup a/b})} \right|^2  . \label{Vfinal}
\end{equation}
In figure~\ref{fig_Berche1} we present a plot of this function for a generic rectangular lattice and for the particular case of a square lattice.

\subsection{Triangular lattice}

For the triangular lattice we consider $\vec{a}_1 =a\hat{x}$ and $\vec{a}_2 = a \cos (\piup/3)\hat{x} +a \sin(\piup/3)\hat{y}$ such that a  point of the lattice located at $\vec{R}_{mn}=m\vec{a}_1+n\vec{a}_2$ will lead to
\begin{equation}
|\vec{r} - \vec{R}_{mn}|^2 = (\varphi - \eta)(\bar{\varphi}-\bar{\eta})\,,
\end{equation}
where $\varphi$ is given by equation~(\ref{phi}) and 
\begin{equation}
\eta = \left(m + \re^{\ri\piup/3}n\right)a .
\end{equation}

Following the steps of the previous section,
\begin{align}
V(x,y) &= \lambda\sum_{(m,n)\in\mathbb{Z}^2} \ln \left[ (\varphi - \eta)(\bar{\varphi} - \bar{\eta}) \right]  
= \lambda\sum_{m=-\infty}^{\infty}  \ln \left[ (\varphi -ma)(\bar{\varphi}-ma) \right]  \nonumber \\ 
&+ \lambda \sum_{m=-\infty}^{\infty}  \sum_{n=1}^{\infty} \ln [ (\varphi^2 - 2ma\varphi +m^2 a^2  + e^{-i\piup/3}n^2 a^2  )\nonumber\\
&\times (\bar{\varphi}^2 - 2ma\bar{\varphi} +m^2 a^2 + \re^{\ri\piup/3}n^2 a^2  ) ] . 
\label{sumtri}
\end{align}

The analogue of equation~(\ref{sum2}) is then
\begin{align}
V(x,y) &= 
\frac{\lambda}{2} \sum_{(m,n)\in\mathbb{Z}^2} \ln  \left( m^2a^2+ \re^{\ri\piup/3}n^2a^2 \right)
+\frac{\lambda}{2} \sum_{(m,n)\in\mathbb{Z}^2} \ln  \left( m^2a^2+ \re^{-\ri\piup/3}n^2a^2 \right) \nonumber \\
&+  \lambda\sum_{m=-\infty}^{\infty} \ln \left\lbrace ma\frac{ \sinh[\piup (\varphi -ma) \re^{\ri\piup/6}/a]}{\sinh (\piup m \re^{\ri\piup/6})}  \right\rbrace 
\nonumber \\
&+  \lambda\sum_{m=-\infty}^{\infty} \ln \left\lbrace ma \frac{ \sinh[\piup(\bar{\varphi} -ma)\re^{-\ri\piup/6} /a]} {\sinh (\piup m \re^{-\ri\piup/6})} \right\rbrace . \label{sum2tri}
\end{align}

After discarding the additive constants, the above expression becomes
\begin{align}
V(x,y)  &=\lambda  \ln[ \sinh(\piup\varphi \re^{\ri\piup/6}/a) \sinh(\piup\bar{\varphi} \re^{-\ri\piup/6}/a)] \nonumber \\
&+ \lambda  \sum_{m=1}^{\infty} \ln \left\lbrace \frac{  \sinh[\piup (\varphi -ma) \re^{\ri\piup/6}/a] \sinh[\piup (\varphi +ma) \re^{\ri\piup/6}/a]}{\sinh^2 (\piup m \re^{\ri\piup/6})} \right\rbrace \nonumber \\
&+ \lambda \sum_{m=1}^{\infty} \ln \left\lbrace  \frac{ \sinh[\piup (\bar{\varphi} -ma) \re^{-\ri\piup/6} /a] \sinh[\piup (\bar{\varphi} +ma) \re^{-\ri\piup/6}/a]}{\sinh^2 (\piup m \re^{-\ri\piup/6})}  \right\rbrace , \label{sum3tri}
\end{align}
in analogy with equation~(\ref{sum3}).

\begin{figure}[!t]
\begin{center}
\includegraphics[width=10cm,trim={0 10cm 3cm 2cm},clip]{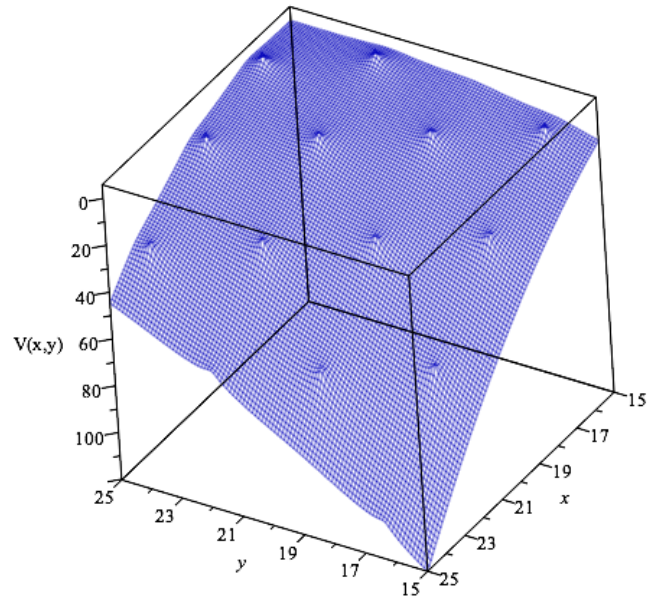}
\caption{(Colour online) Three-dimensional plot of the ``potential function'' $ V(x,y)$ for the triangular lattice.}
\label{fig_Berche2}
\end{center}
\end{figure}

In terms of the coordinates $x$ and $y$, the final expression for the regularized potential is then
\begin{align}
V(x,y) = \lambda \ln \left\lbrace  \frac{\vartheta_1 \left[\ri\frac{\piup}{a}\left(x+\ri y\right) \re^{\ri\piup/6}, -\ri \re^{-\sqrt{3}\piup/2 }\right]   \cdot  \vartheta_1 \left[\ri\frac{\piup}{a}\left(x-\ri y\right) \re^{-\ri\piup/6},\ri\re^{-\sqrt{3}\piup/2 }\right]}{\cosh^2\left[\frac{\piup}{2a}\left(y-\sqrt{3}x\right)\right]-\cos^2\left[\frac{\piup}{2a}\left(x+\sqrt{3}y\right)\right]}  \right\rbrace .  
\label{Vregtrig} 
\end{align} 
A graphic representation of this function can be seen in figure~\ref{fig_Berche2}.

\begin{figure}[!t]
\begin{center}
\includegraphics[width=8cm]{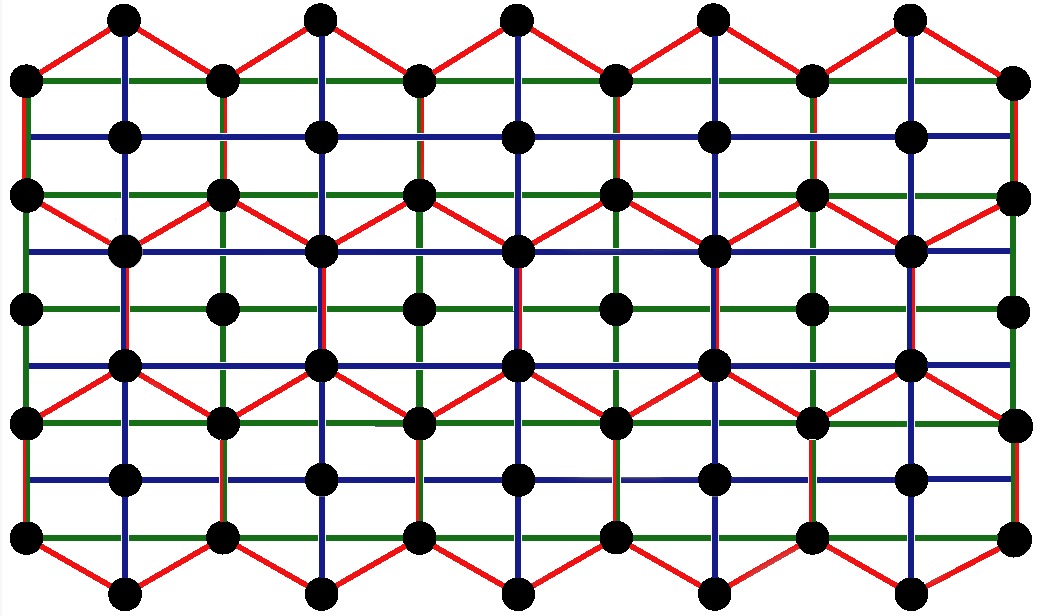}
\caption{(Colour online) Representation of the triangular lattice as two  rectangular lattices shifted with respect to each other.}
\label{fig_2rectangular}
\end{center} 
\end{figure}

Due to the linearity of equation~(\ref{laplacian}), the above result can also be obtained from the superposition of the potentials of two rectangular lattices displaced  relatively to each other in such a way as to form the triangular lattice  (see figure~\ref{fig_2rectangular}).

\section{Conclusion}

{In this paper, we performed  infinite logarithmic sums, with proper regularization, to determine the interaction energy of rectangular and triangular lattices of line defects having their axes along the $z$-direction. By adjusting the defect strength $\lambda$,  along with parameters governing the geometry of a cell (namely $a$, $b$), one has the possibility to perform defect engineering, that is tailoring material properties from controlled defect arrays \cite{Fumeron2017,0957-4484-27-13-135302}.}

Particles moving inside a lattice of topological defects may be highly sensitive to initial conditions and hence the dynamics of these particles is likely to lead to exponential divergence of initially closed trajectories. The motion of fast electrons in a silicon crystal endowed with periodically distributed atomic strings is known to be chaotic \cite{Shulga} and deserves a separate treatment involving the statistical tools of dynamic hamiltonian systems. This will be the object of a next study. 


\section*{Acknowledgements}
F.M. is thankful for the financial support and warm hospitality  of the Statistical Physics Group at Universit\'e de Lorraine. This work was partially funded by INCT nanocarbono, CNPq, CAPES and FACEPE (Brazilian agencies) and PICS CNRS (France). We also thank Eudes Gomes for helping with figure~\ref{fig_2rectangular}.


\ukrainianpart

\title{Стосовно енергії граток з топологічними дефектами}
\author{Б. Берш\refaddr{label1}, С. Фумерон\refaddr{label1}, Ф. Мораес\refaddr{label2}}
\addresses{
	\addr{label1} Динаміка і симетрія, лабораторія фізики і теоретичної хімії,  CNRS - університет Лоррен, UMR 7019, Вандувр лє Нансі, Франція
	\addr{label2} Кафедра фізики, Федеральний  університет Пернамбуку
	52171--900 Ресіфі, Бразилія}

\makeukrtitle 

\begin{abstract}
	Оскільки логарифмічна функція є розв'язком рівняння Пуасона у двох вимірах, вона фактично є кулонівською взаємодією у двовимірному випадку, взаємодією між лініями потоку Абрикосова  у надпровіднику   II типу, або між лінійними дефектами у пружних середовищах, і т. п. Ось чому гратки ліній, що взаємодіють логарифмічно є предметом інтенсивних досліджень завдяки їх багатогранним застосуванням. Ще з кінця 1990-х років розв'язок рівняння Пуасона для такиз граток був відомий у вигляді нескінченої суми. У даній статті представлено альтернативний аналітичний розв'язок у замкнутому вигляді в термінах тета-функції Якобі.
	\keywords топологічний дефект, космічна струна, лінія потоку
	
\end{abstract}

\lastpage
\end{document}